\documentclass[fleqn]{2020SCGE}
\usepackage{color}
\usepackage{float}
\usepackage{gensymb}
\usepackage{graphicx}
\usepackage{tablefootnote}
\usepackage[toc]{multitoc}
\usepackage{rotating}
\usepackage{lscape}
\usepackage{pdflscape}
\usepackage{setspace}
\usepackage[numbers]{natbib}
\usepackage{lineno}
\usepackage{hyperref}
\usepackage{graphicx}	
\usepackage{threeparttable}

\usepackage{threeparttable}
\usepackage{color}
\usepackage[dvipsnames]{xcolor}
\usepackage[normalem]{ulem}
\usepackage{booktabs}
\usepackage{multirow}
\def \link_col{blue}

\def\gray{$\gamma$-ray\ }
\def\grays{$\gamma$-rays\ }
\def\fermi{{\it Fermi }}

\usepackage{url}
\usepackage{hyperref}
\usepackage[utf8]{inputenc}
\usepackage{amsmath}	

\newcommand{\aap}{Astron. \& Astrophys.}

\newcommand{\apjl}{Astrophys. J. Lett.}
\newcommand{\apjs}{Astrophys. J. Suppl.}

\let\oldequation\equation
\let\oldendequation\endequation
\renewenvironment{equation}
{\linenomathNonumbers\oldequation}
{\oldendequation\endlinenomath}
\begin{document}

\def\bm{\boldsymbol}

\def\dl{\displaystyle}
\def\du{\end{document}}
\def\d{{\rm d}}
\def\e{{\rm e}}
\def\i{{\rm i}}

\ensubject{subject}


\ArticleType{Article}
\SpecialTopic{SPECIAL TOPIC: }
\Year{2024}
\Month{?}
\Vol{?}
\No{?}
\DOI{??}
\ArtNo{000000}
\ReceiveDate{?}
\AcceptDate{?}
\title[LHAASO J0248+6021]{LHAASO detection of very-high-energy gamma-ray emission surrounding PSR J0248+6021}

\author{LHAASO Collaboration\footnote{Corresponding authors: caowy@mail.ustc.edu.cn, lbing@ustc.edu.cn, yangrz@ustc.edu.cn, yuyh@ustc.edu.cn} \\(The LHAASO Collaboration authors and affiliations are listed after the references.)}{}%

\AuthorMark{Zhen Cao}

\AuthorCitation{Zhen Cao, et al}

\abstract{We report the detection of an extended very-high-energy (VHE) gamma-ray source coincident with the location of middle-aged (62.4~\rm kyr) pulsar PSR J0248+6021, by using the LHAASO-WCDA data of live 796 days and LHAASO-KM2A data of live 1216 days. A significant excess of \gray induced showers is observed both by WCDA in energy bands of 1-25~\rm TeV and KM2A in energy bands of $>$ 25~\rm TeV with 7.3 $\sigma$ and 13.5 $\sigma$, respectively. The best-fit position derived through WCDA data is R.A. = 42.06$^\circ \pm$ 0.12$^\circ$ and Dec. = 60.24$^\circ \pm $ 0.13$^\circ$ with an extension of 0.69$^\circ\pm$0.15$^\circ$ and that of the KM2A data is R.A.= 42.29$^\circ \pm $ 0.13$^\circ$ and Dec. = 60.38$^\circ \pm$ 0.07$^\circ$ with an extension of 0.37$^\circ\pm$0.07$^\circ$. No clear extended multiwavelength counterpart of this LHAASO source has been found from the radio band to the GeV band. The most plausible explanation of the VHE \gray emission is the inverse Compton process of highly relativistic electrons and positrons injected by the pulsar. These electrons/positrons are hypothesized to be either confined within the pulsar wind nebula or to have already escaped into the interstellar medium, forming a pulsar halo.  
}

\keywords{Key Words:~~\textnormal{gamma-rays, pulsars: individual: PSR J0248+6021, Interstellar medium (ISM), nebulae}}
\PACS{95.80.+p, 98.62.Py, 98.52.-b, 95.55.Jz, 95.30.Ky, 98.58.Ge}

\maketitle

\begin{multicols}{2}
\section{Introduction} \label{sec:intro}

 Pulsars are one of the most energetic source populations in our Galaxy. They are bright emitters in both radio and GeV \gray wavelengths. The relativistic wind generated by pulsars can interact with the surrounding medium to form a structure known as a pulsar wind nebula (PWN), which has been detected across the entire electromagnetic spectrum. Furthermore, HAWC collaborations have detected very extended VHE \gray structures from two middle-aged pulsars Geminga and Monogem \citep{hawc_geminga}. The radial profiles of these structures are consistent with the predicted inverse Compton emission from electrons injected by the PWN as they propagate through the interstellar medium. Such a structure is known as a pulsar halo. Another interesting property is that the diffusion coefficient within these structures is much smaller than the average diffusion coefficient in the Galactic plane, which can have significant implications for cosmic ray (CR) propagation and Galactic diffuse \gray emissions (GDE) \citep{ryliu2020, Yan2023PhRvD,scpma}.  Generally, PWNe are formed beyond the termination shock (TS) of the pulsar wind and are dominated by relativistic electron-positron pairs and
magnetic fields. The particle transport is dominated by advection. While at the later stage of the PWNe evolution ($t > 10^5 $ yrs) the electron/positrons accelerated in the TS eventually escaped diffusively to the ambient ISM and formed more extended halo structures \citep{giacintiHaloFractionTeVbright2020b, lopez-coto22, linden19}. The study on the \gray emissions in these structures would have important implications on the particle acceleration and propagation in the vicinity of pulsars \citep{lopez-coto22}.

PSR J0248+6021, with a rotation period of $P=217$~ms and a spin-down power of $L_{\rm sd}=2.13\times10^{35}\,{\rm erg\, s}^{-1}$, was first discovered in a survey conducted by the Nancay radio telescope \citep{Foster1997}. It was later detected in the GeV band by the Fermi Large Area Telescope (LAT) \citep{Abdo2010b}. According to Theureau et al.\citep{Theureau2011}, the PSR J0248+6021 is likely located within a dense environment, specifically the giant HII region W5 at a distance of 2 kpc. However, no significant X-ray counterpart or pulsar wind nebula (PWN) associated with PSR J0248+6021 has been detected yet \citep{Mignani2016}.

In this paper, we report the detection of an extended \gray source from the Large High-Altitude Air Shower Observatory (LHAASO), which is potentially a pulsar halo associated with PSR J0248+6021. The paper is organized as follows: In Sec.\ref{sec:lhaaso}, we present the analysis results of both KM2A and WCDA data. In Sec.\ref{sec:multi}, we explore the multiwavelength studies of the newly discovered LHAASO source, focusing on the Fermi-LAT GeV observations and gas distributions. Finally,  we discuss the possible origin of this source in Sec.\ref{sec:dis}.

\section{LHAASO Data analysis} 
\label{sec:lhaaso}
\subsection{Analysis method} 
LHAASO, a hybrid observatory, is designed for the study of cosmic rays and gamma rays across a broad energy range, from hundreds of GeV to PeV. It is composed of three sub-arrays, namely, the Water Cherenkov Detector Array (WCDA), the Kilometer Squared Array (KM2A), and the Wide Field-of-view air Cherenkov/fluorescence Telescope Array (WFCTA). The WCDA, covering an area of 78,000 m$^2$, is dedicated to TeV \gray astronomy. The KM2A, comprising 5,195 electromagnetic particle detectors (EDs) and 1,188 muon detectors (MDs), is utilized for \gray astronomy above 10~\rm TeV. More details about the detectors and the reconstruction methods of WCDA and KM2A can be found in \citep{aharonianPerformanceLHAASOWCDAObservation2021a,aharonianObservationCrabNebula2021}. The data used in this work for the WCDA were collected from March 5, 2021, to July 31, 2023. For KM2A, data were obtained using the half array, the quarter array, and the full array from December 27, 2019, to July 31, 2023. After the data quality check, the number of events used in this analysis is 4.69$\times$10$^9$ for WCDA and 1.70$\times$10$^9$ for KM2A.

The WCDA uses the number of triggered PMT units, referred to as $N_{\rm hits}$, as the shower energy estimator, and events are divided into six groups, i.e., 60--100, 100--200, 200--300, 300--500, 500--800, $>$ 800. The KM2A data sets are divided into five groups per decade with a bin width of $\Delta\log_{10}$E = 0.2 According to the reconstructed energy. The sky map in the celestial coordinate system (right ascension and declination) is divided into a grid of 0.1$^\circ\times$0.1$^\circ$ and each cell is filled with the number of the detected events According to their reconstructed arrival direction. The "direct integration method" \citep{Fleysher_2004} is adopted to estimate the number of CR background events in each cell.

A 3-dimensional (3D) likelihood algorithm was used to 
fit the morphology and spectrum of the source simultaneously, and the test statistic (TS) was used to evaluate the significance of the source. TS value is defined as 2log($\mathcal{L}_{1}$/$\mathcal{L}_{0}$), where $\mathcal{L}_{1}$ is the maximum likelihood value for the alternative hypothesis and $\mathcal{L}_{0}$ is the maximum likelihood value of the null hypothesis.  According to Wilks' Theorem \citep{wilks1938}, the TS value follows a chi-square distribution with the number of free parameters in the signal model. In this work, to estimate the significance of the sky map, we assume the source is a point source with a power-law spectrum. For WCDA, the spectrum has an index of 2.6 in the energy range 1-25~\rm TeV, and for KM2A, the spectrum has an index of 3.0 at energies greater than 25~\rm TeV. In each pixel, the flux is the only variable parameter, and According to Wilks' Theorem, we take $\pm\sqrt{\rm TS}$ as the significance.  

The detection of source components is an iterative process. To implement the fitting progress, a region of interest (ROI) with a radius of 6$^\circ$ centered at the right ascension (R.A.) 42.0$^\circ$ and declination (Dec.) 60.5$^\circ$ is used. $\Delta TS$, defined as 2ln($\mathcal{L}_{N+1}/\mathcal{L}_{N}$), where $\mathcal{L}_{N}$ and $\mathcal{L}_{N+1}$ represents the maximum likelihood of model with $N$ and $N +1$ sources,  respectively, is used to compare models with $N$ and $N+1$ source components. An additional source will be accepted in this ROI when $\Delta TS >$ 25. To account for the GDE, a diffuse source with a simple power law spectrum and a spatial distribution derived from the dust optical depth measured by Planck telescope\citep{refId0,refId1} is added during the source search process.

\subsection{Results}
\begin{figure*}
    \centering
     \includegraphics[width=0.85\linewidth]{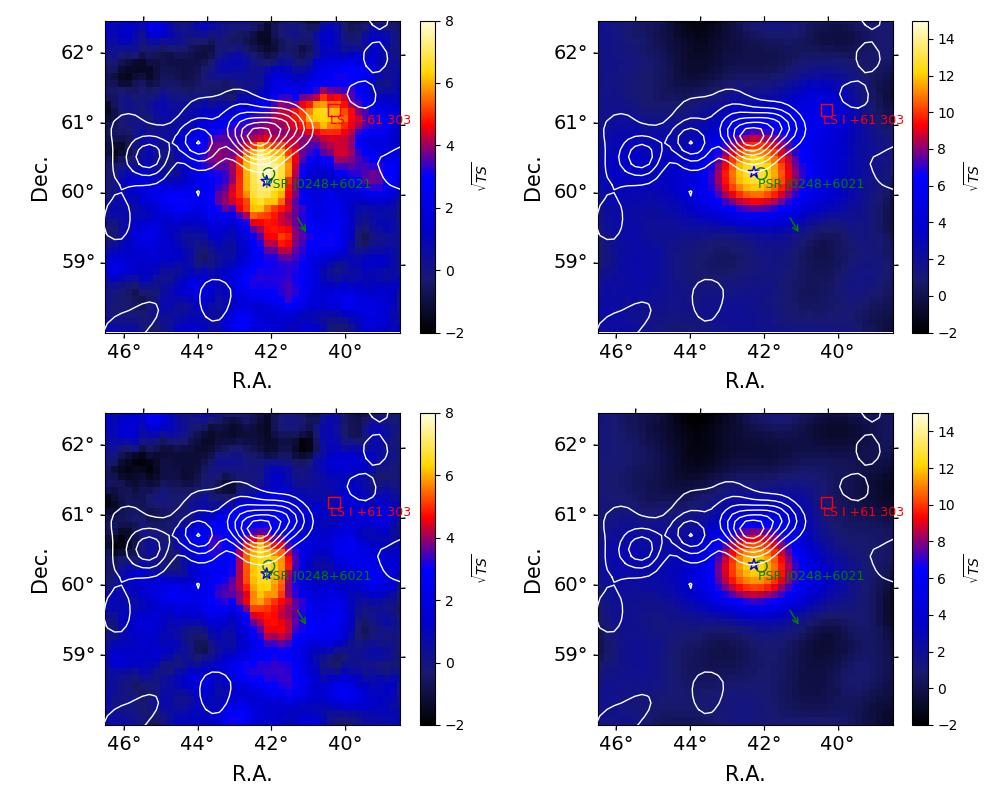} 
    \caption{The top left panel: The WCDA significance map at energy 1--25~\rm TeV. The top right panel: The KM2A significance map above 25~\rm TeV. The bottom left panel: The WCDA significance map at energy 1--25~\rm TeV after subtracting all sources except LHAASO J0248+6021. The bottom right panel: The KM2A significance above 25~\rm TeV after subtracting all sources except LHAASO J0248+6021. The blue star denotes the best-fit position of the LHAASO source. The green circle shows the location of PSR J0248+6021, and the green arrow shows the possible birth position of the pulsar given the direction and velocity of its transverse motion  $v_{\rm T}\sim 500\, {\rm km\,s}^{-1}$ \citep{Theureau2011}. The red square marks the location of LS I +61 303. The CO data has been utilized to derive the gas column density distribution, illustrated by the white contours, with values exceeding $1 \times 10^{21}~\rm cm^{-2}$. This extraction was performed over an integrated velocity range ranging from -30 ~\rm km/s to -50~\rm km/s. }
    \label{fig:Figure1}
\end{figure*}

Three sources were identified through iterative searches within the ROI, two of which are close to PSR J0248+6021 based on their angular distance. One is an extended source that is consistent with 1LHAASO J0249+6022 from the first LHAASO catalog \citep{lhaaso_catalog_apjs}. The other is a point source with an angular distance of 1.20$^\circ$ from the pulsar, which is very close to the gamma-ray binary LS I +61 303. The TS of the GDE in this ROI, derived through WCDA data, is 4.0, while the TS derived through KM2A data is 10.2. The GDE flux of this region is consistent with the flux of the outer Galaxy region as shown in Cao et al.\citep{KM2A_Diffuse_prl}.
In this work, we mainly focused on the VHE \grays around the pulsar ( 1LHAASO J0249+6022, hereafter referred to as LHAASO J0248+6021), and other sources were treated as background. Fig.~\ref{fig:Figure1} shows the smoothed significance maps around PSR J0248+6021 before and after subtracting other background sources. The max significance of LHAASO J0248+6021 is 7.3 $\sigma$ at energy range 1--25~\rm TeV and 13.5 $\sigma$ at energy greater than  25~\rm TeV.  

We employed a two-dimensional Gaussian model and a continuous injection diffusion model from a point source as described in Eq.~\ref{Equ: Equ1} \citep{tangPositronFluxGray2019,dimauroDetectionRayHalo2019,2021PhRvL.126x1103A,scpma, hawc_geminga} to study the morphology of this extended source.  In Eq.~\ref{Equ: Equ1}, $\theta$ is the angular distance from the source position, and $\theta_{d}$ is the diffusion extension.  The spectrum of LHAASO J0248+6021 is assumed as a power law $f$(E) = $J\times(E/E_{0})^{-\alpha}$. The reference energy $E_{0}$ is set as 40~\rm TeV for KM2A, and 3~\rm TeV for WCDA, respectively. We found that the continuous injection diffusion model yielded a better Akaike Information Criterion \citep{Akaike1974} (AIC) value with the same number of free parameters as the Gaussian model. Compared to the Gaussian model, the AIC value for the diffusion model decreased by 6.8 for WCDA and 6.3 for KM2A.  Additionally, we used an asymmetric Gaussian model for the WCDA data, which appears to exhibit an asymmetric morphology. Compared to the symmetric Gaussian model, the TS value for the asymmetric Gaussian model increased by 5.74 with two additional free parameters. However, this improvement is not significant based on the current analysis. Notably, the major axis determined by the fitting process is closely aligned with the direction of the proper motion of PSR J0248+6021, as shown in Fig.~\ref{fig:Figure1}. 


\begin{equation} \label{Equ: Equ1}
    f(\theta)\propto\frac{1}{\theta_{d}(\theta+0.085\theta_{d})} \exp\left[-1.54\left(\frac{\theta}{\theta_{d}}\right)^{1.52}\right]
\end{equation}

\begin{figure*}
    \centering
    \includegraphics[width=0.9\linewidth]{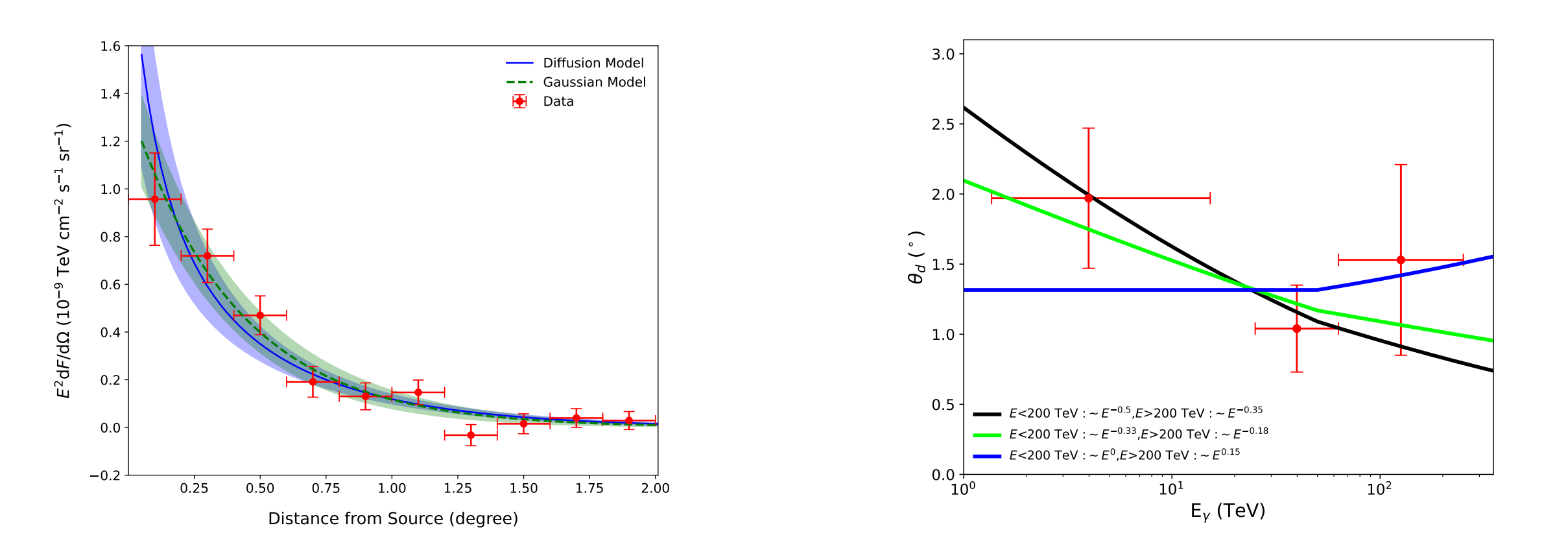}
    \caption{Left panel: One-dimensional distribution of $\gamma$ ray flux of LHAASO J0248+6021. The  blue (green) solid line represents the best-fitting model and the blue (green) shaded band is the $\pm$1$\sigma$ statistical uncertainty, which is the convolution of Eq.~\ref{Equ: Equ1} (Gaussian model) with the PSF. Right panel: The $\theta_{d}$ varies with the energy. The E in the legend represents the energy of the electrons.  The result of the energy bin 1--25~\rm TeV is derived from WCDA data, and those of energy bins 25--63~\rm TeV and 63--251~\rm TeV are derived from KM2A data, respectively.}
    \label{fig:Figure2}
\end{figure*}

In the case of the diffusion model, the best-fit position derived from WCDA data is R.A. = 42.11 $^\circ\pm$ 0.16 $^\circ$ and Dec. = 60.28$^\circ\pm$ 0.11$^\circ$, while for KM2A data, it is R.A.= 42.29 $^\circ\pm$ 0.13$^\circ$ and Dec. = 60.38 $^\circ\pm$ 0.07$^\circ$. The centroid of the LHAASO source is consistent with the location of PSR J0248+6021, with a distance of 0.11$^\circ\pm$0.20$^\circ$ for WCDA and 0.11$^\circ\pm$0.15$^\circ$ for KM2A. The fitted $\theta_{d}$ is 1.97$^\circ\pm$0.5$^\circ$ in the energy range of 1-25~\rm TeV and 1.11$^\circ\pm$0.28$^\circ$ for $E >$ 25~\rm TeV. The details of fitting results are shown in Tab.~\ref{tab:tab1}. The left panel of Fig.~\ref{fig:Figure2} shows the one-dimensional distribution of the $E >$ 25$~\rm$ TeV \gray emission of LHAASO J0248+6021 after subtracting the estimated other sources, along with fitting 1 $\sigma$ band of the diffusion model.

To further study the energy dependence of the spatial distribution of the source, we derived the $\theta_{d}$ in three energy bins, 1--25~\rm TeV derived through the WCDA data, 25--63~\rm TeV, and 63--251~\rm TeV derived through the KM2A data, respectively. To get the $\theta_d$, We fix the index of the spectrum as the best-fit value by fitting the data with $E>$ 25~\rm TeV. The results are shown in the right panel of Fig.~\ref{fig:Figure2}. Due to the limited statistics, the current data are more or less consistent with an energy-independent behavior, but there are hints of the decrease in size from the first energy bin to the second energy bin.  

\begin{figure*}
    \centering
    \includegraphics[width=0.9\linewidth]{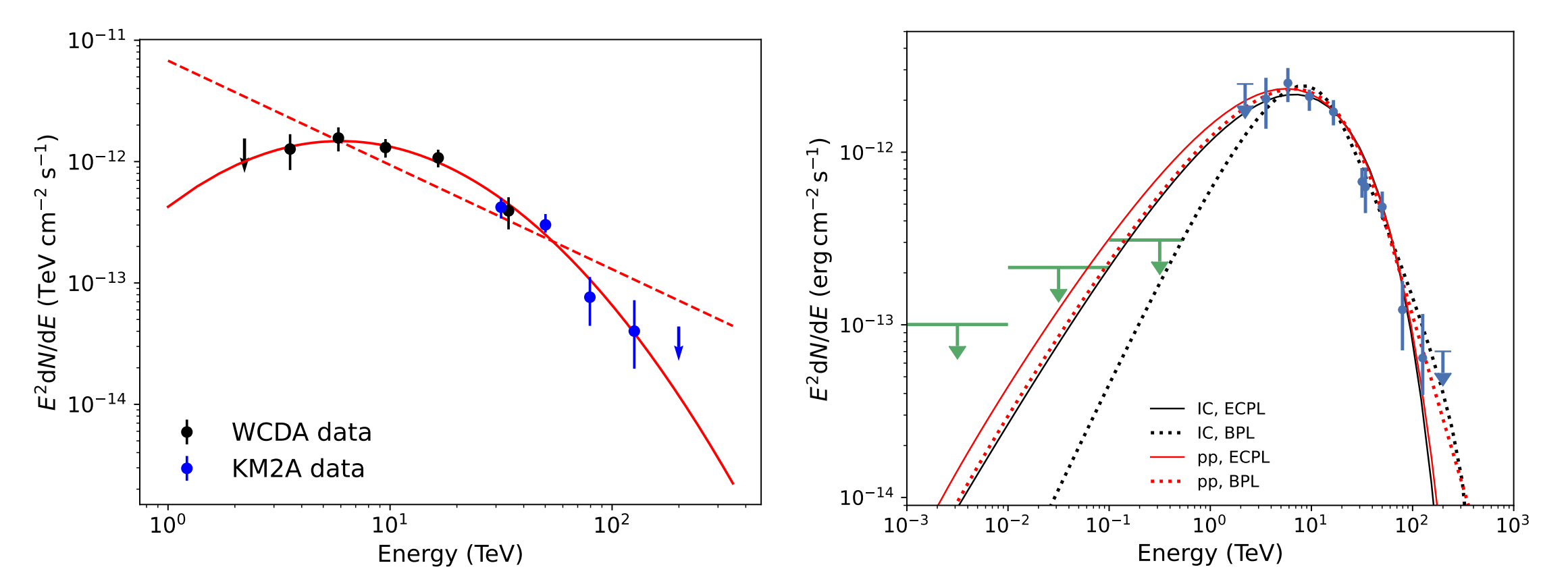} 
    \caption{Left panel: The energy spectrum of LHAASO J0248+6021. The solid line is the best-fit result assuming a log-parabola function, and the dotted line is the result of a single power-law. Right panel: SED fit results for different radiation models of LHAASO J0248+6021, and the green data ( $95\%$ upper limits are derived from Fermi-LAT observations. }
    \label{fig:Figure3}
\end{figure*}

Using the diffusion model with $\theta_d$ of 1.97$^\circ$ for WCDA and 1.11$^\circ$ for KM2A, and assuming a single power-law, the differential flux (TeV$^{-1}$cm$^{-2}$s$^{-1}$) derived through the WCDA data in the energy range from 1 TeV to 25 TeV is (1.44$\pm$0.44)$\times$10$^{-13}$($E$/3~\rm TeV)$^{-2.43\pm0.13}$. The differential flux derived through the KM2A data in the $E >$ 25~\rm TeV is (2.24$\pm$0.49)$\times$10$^{-16}$($E$/40~\rm TeV)$^{-3.84\pm0.25}$. The spectral energy distribution (SED) is shown in Fig.~\ref{fig:Figure3}. We used a power-law spectrum $f(E) = J\times(E/E_{0})^{-\alpha}$, and the log-parabola spectrum $f(E) = J\times(E/E_{0})^{-(\alpha+\beta ln(E/E_{0}))}$, to fit the WCDA and KM2A data simultaneously. The reference energy $E_{0}$ is chosen to be 30~\rm TeV here. The obtained best-fit parameters are $J$=(3.67$\pm$0.31)$\times$10$^{-16}$~\rm TeV$^{-1}$s$^{-1}$cm$^{-2}$, $\alpha=2.76\pm0.06$ with $\chi^{2}/ndf$=31.2/2 fitting with the power-law function and $J$=(5.40$\pm$0.50)$\times$10$^{-16}$~\rm TeV$^{-1}$s$^{-1}$cm$^{-2}$, $\alpha=3.14\pm0.13$, $\beta=0.39\pm0.09$ with $\chi^{2}/ndf$ = 5.3/3 fitting with the log-parabola function. Compared with a single power-law fit, the improvement is about 5.1 $\sigma$ using a log-parabola function, which reveals a clear spectral curvature in the \gray SEDs. 

The systematic uncertainty has been investigated in Aharonian et al.\citep{aharonianPerformanceLHAASOWCDAObservation2021a,aharonianObservationCrabNebula2021}. It is mainly contributed by the atmospheric model used in the Monte Carlo simulations. The overall systematic uncertainty affecting the KM2A spectrum measurement is estimated to be $\sim$ 7$\%$ on the flux and 0.02 on the spectral index. The total systematic uncertainty affecting the WCDA flux measurement is $\sim  8\%$ \citep{lhaaso_catalog_apjs}.  Additionally, the systematic uncertainties from the measurement of GDE flux should also be considered. The Galactic Longitude of LHAASO J0248+6021 is 137$^\circ$, where the column density is relatively low, and the GDE should only have a minor impact. We also test the LHAASO J0248+6021 SED, assuming that there is no GDE. For WCDA, the GDE can induce an uncertainty of about 26$\%$ on the measured flux, while for KM2A, it is 30$\%$, which is approximately 1.5 times the statistical error. 

\begin{table*}
\centering
\caption {Fitting Results of LHAASO J0248+6021 by WCDA and KM2A.}
\begin{tabular}{lllllll}
\hline 
& & R.A. & Dec. & $\sigma$/$\theta_{d}$ $^{a}$& FLux$^{b}$ & $\alpha$ \\
\hline
\multirow{2}{*}{Gaussian model}   & WCDA & 42.06$\pm$0.12$^\circ$ &60.24$\pm$0.13$^\circ$& 0.69$\pm$0.15$^\circ$  &  1.54$\pm$0.43     &    2.43$\pm$0.15  \\
                                  & KM2A &  42.27$\pm$0.14$^\circ$ & 60.41$\pm$0.07$^\circ$     &  0.37$\pm$0.07$^\circ$     & 1.82$\pm$0.34 &3.86$\pm$0.26    \\
\hline
\multirow{2}{*}{Diffusion model} & WCDA &  42.11$\pm$0.16$^\circ$ &  60.28$\pm$0.11$^\circ$      &  1.97$\pm$0.50$^\circ$    &   1.83$\pm$0.50 &2.41$\pm$0.11\\
                              & KM2A &   42.29$\pm$0.13$^\circ$ & 60.38$\pm$0.07$^\circ$     & 1.11$\pm$0.28$^\circ$  & 2.24$\pm$0.49&3.84$\pm$0.25   \\
\hline 
\end{tabular}

{\footnotesize $a$ , for Gaussian model is $\sigma$ and for diffuse model is $\theta_d$.  \\
 $b$, for WCDA the flux unite is 10$^{-13}$~\rm TeV$^{-1}$s$^{-1}$cm$^{-2}$ and the reference energy is 3~\rm TeV\\
    for KM2A the flux unite is 10$^{-16}$~\rm TeV$^{-1}$s$^{-1}$cm$^{-2}$ and the reference energy is 40~\rm TeV.  }
\label{tab:tab1}
\end{table*}

\section{MULTIWAVELENGTH OBSERVATIONS} 
\label{sec:multi}

\subsection{Fermi-LAT observation}


To have a better understanding of the diffuse \gray emission around the PSR J0248+6021, we analyzed about 15 years of Fermi-LAT Pass 8 data (from August 4, 2008, until September 19, 2023) in this region using the Fermitools from Conda distribution\footnote{\url{https://github.com/fermi-lat/Fermitools-conda/}}. Here, we only selected source class events with energy above 1 GeV to have a better angular resolution, chose a $20^{\circ}\times20^{\circ}$ region centered at LHAASO J0248+6021 as the ROI, and performed the standard binned Likelihood analysis to search for possible GeV gamma-rays that related to the LHAASO source.  We used the make4FGLxml.py to generate the background model based on the newly updated LAT 14-year source catalog \citep[4FGL-DR4,][]{balletFermiLargeArea2023}. The background model includes the point source 4FGL J0248.4+6021 associated with pulsar PSR J0248+6021, other GeV sources with angular distances to the center of ROI less than $20^{\circ}$, the Galactic diffuse emission (gll\_iem\_v07.fits), and isotropic diffuse emission (iso\_P8R3\_SOURCE\_V3\_v1.txt).  The spectral models of the above sources are all adopted from 4FGL-DR4.  
We then applied the maximum likelihood fitting to optimize the spectral parameters of the sources in the background model and refer to the fitted results as the best-fit background model. 
Using the best-fit background model, we generated a residual TS map around PSR J0248+6021 (as shown in Fig.~\ref{fig:Figure4}) and found no significant GeV emission left after subtracting the \gray emission from the pulsar and other background sources. 
Therefore, to obtain a constraint on the GeV band flux of the LHAASO detected very high energy diffuse emission around PSR J0248+6021, we then added an extended source applying the same diffusion model acquired from KM2A data analysis to the Fermi-LAT source model and assuming its spectral shape is a simple power-law. We then performed a new likelihood analysis using the new source model and found the significance of the added extended source is lower than 2\,$\sigma$.  
Finally, to obtain the SED of J0248+6021 in the GeV range, we performed the maximum likelihood analysis in three logarithmically spaced energy bins for events of 1--1000 GeV.  For each energy bin, the power-law index of  the added extended source is fixed to be 1.62, which is the best-fitting result for the Fermi-LAT data (1-1000 GeV). The significance of J0248+6021 of each bin is lower than 1\,$\sigma$, thus we derived the $95\%$ upper limits of the fluxes, and the results are shown in Fig.~\ref{fig:Figure3}. 
In addition, we also tested the diffuse model acquired from WCDA data analysis, the GeV flux of the source is basically the same as that of the model acquired from KM2A data analysis.

\subsection{Gas distribution and other possible counterparts}

 To find whether the origin of the LHAASO detected diffuse \grays are hadronic, i.e., the \grays are produced via the proton-proton inelastic collision among CR nuclei and the gases, we also explored the gas distribution along the line of sight. First, we generated the total CO intensity ($W_{\rm CO}$ in unit of ${\rm K}\,{\rm km\,s}^{-1}$) map integrated over the velocity range $-110$--$+110\,{\rm km\,s}^{-1}$ using CO observation data from Dame et al.\citep{Dame2001}.  As shown in the right panel of Fig.\ref{fig:Figure4}, there are dense molecular gas partially overlapped with the LHAASO source along the line of sight.  
Next, to find which specific cloud is more likely associated with LHAASO J0248+6021, we generated CO intensity maps integrated over successive 20\,${\rm km\,s}^{-1}$ intervals in the velocity range $-110$ - $+110\,{\rm km\,s}^{-1}$. By comparison, we found the dense molecular cloud that partially overlapped with the UHE \grays are mainly within the velocity range of $-50\,{\rm km\,s}^{-1}$--$-30\,{\rm km\,s}^{-1}$. Moreover, these molecular gases are associated with the giant star-forming region W5 (also known as the Soul nebula) at a distance about 2 kpc, in which the ionized gases are within the similar velocity range (from $-49\,{\rm km\,s}^{-1}$ to $-31\,{\rm km\,s}^{-1}$) \citep[see,e.g.,][]{Lada1978,Heyer1998,Karr2003,Theureau2011}.

\section{Discussion and conclusion} 
\label{sec:dis}
  
To investigate the possible radiation mechanisms of the \grays in this region, we fit the SEDs acquired from  Fermi-LAT, WCDA, and KM2A  observation with both leptonic scenario, i.e., the inverse Compton scattering (hereafter referred to as IC model) and hadronic scenario, i.e., proton-proton inelastic collision (hereafter referred as PP model). The fitting was performed using the Naima package\footnote{\url{http://naima.readthedocs.org/en/latest/index.html}} \citep{naima}, which includes tools to perform Markov Chain Monte Carlo fitting of nonthermal radiative processes to the data and allows us to implement different functions. Here, the distribution function of the parent particles was assumed to be ExponentialCutoffPowerLaw (ECPL) or BrokenPowerLaw (BPL)(see Tab.~\ref{tab:form} for the formulae of these distribution functions). The best-fit results of the spectral parameters and the maximum Log(likelihood) (MLL) of each model are presented in Tab.~\ref{tab:sedfit}.  Meanwhile, the resulting spectra are illustrated in Fig.~\ref{fig:Figure3}, in which the black lines are the results of IC model fitting and the red lines are the results of PP model fitting. The seed photon field for relativistic electrons to scatter only includes the cosmic microwave background (CMB).

\begin{figure*}
\centering
  \includegraphics[width=0.85\textwidth]{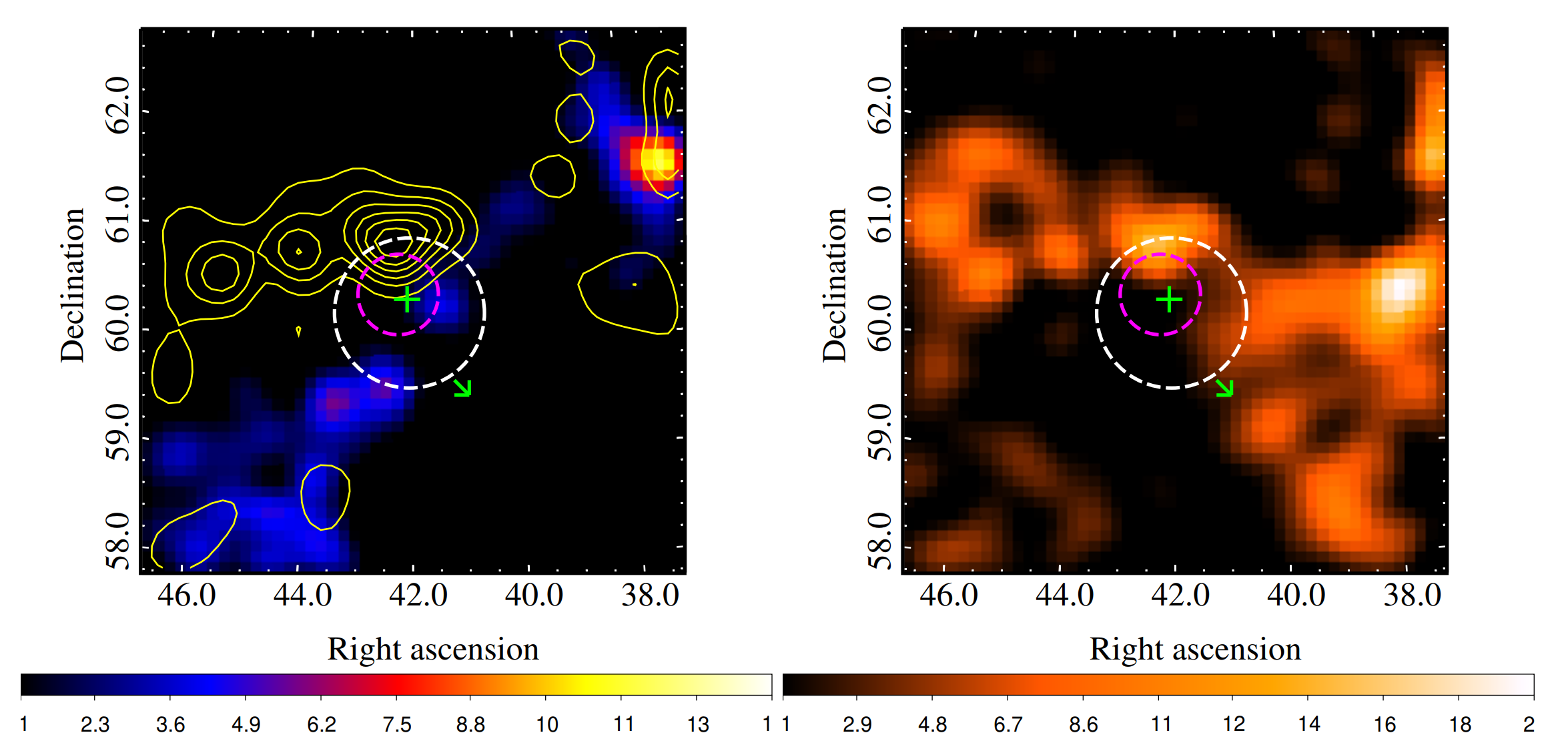}
\caption{ Left panel: The GeV residual TS map around PSR J0248+6021 in which the contribution of sources in the 4FGL-DR4 catalog are subtracted as background. Right panel: CO intensity map integrated over from $-110$ to $110\,{\rm km\,s}^{-1}$ using survey data from  Dame et al.\citep{Dame2001} (in unit of ${\rm K}\,{\rm km\,s}^{-1}$).  The green cross shows the position of PSR J0248+6021, and the green arrow shows the possible birth position of the pulsar given the direction and velocity of its transverse motion  $v_{\rm T}\sim 500\, {\rm km\,s}^{-1}$ \citep{Theureau2011}. The dashed circle with radii of $\sigma$ (the Gaussian disk models) obtained from WCDA data is shown in white and that from KM2A data is shown in magenta.  The yellow contours show the CO intensity distribution integrated over the velocity range of $-50\,{\rm km\,s}^{-1}$ to $-30\,{\rm km\,s}^{-1}$ ($W_{\rm CO}>2\,{\rm K}\,{\rm km\,s}^{-1}$) using data from  Dame et al. \citep{Dame2001}.}
\label{fig:Figure4}
\end{figure*}

As shown in Fig.~\ref{fig:Figure3} and Tab.~\ref{tab:sedfit}, both PP and IC models can fit the SEDs data well, and the acquired maximum likelihood values for BPL spectra are basically the same. To produce the diffuse \gray emission, the required energy budget for electrons  $W_e$ is $\sim5\times10^{45}$\,erg, and budget for the protons $W_p$ is $\sim9/(n_{\rm H}/1 {\rm cm}^{-3})\times10^{48}$\,erg, in which $n_{\rm H}$ is the number density of the atomic hydrogen in the medium. Both are within a reasonable range, which can be provided by the pulsar or a typical supernova remnant. Young stellar clusters are also candidate CR accelerators that have been detected via $\gamma$-ray observations \citep[e.g.,][]{m17,rosette,2019Felixyang}. Thus, the open cluster in W5, i.e., IC 1848, may be responsible for the VHE \grays. Moreover, the gases around this region, as shown in Fig.~\ref{fig:Figure4}, can provide targets for CR protons accelerated by the cluster.  Therefore, the hadronic origin is possible although there is no SNR found around the detected high-energy $\gamma$-rays. Moreover, an undetected SNR may also be the power source of this \gray source if only the required energy budget ($W_p$ or $W_e$) is considered. 
 As shown in Tab.~\ref{tab:sedfit}, the spectral index of the high-energy protons required to generate the detected \grays via the pion-decay process is about  $-1$ or even harder. Such a hard proton spectrum can hardly be realized in shock acceleration in SNRs, but the hard spectrum may be caused by transport effects assuming the \gray emission is from molecular clouds illuminated by propagated CRs.  However, the distribution of the \grays is only partially overlapped with the molecular gas and the best-fit position of the \gray emission is much closer to the pulsar instead of the dense core of the molecular gas (as shown in Fig.\ref{fig:Figure1}).
Meanwhile, the hard spectrum with a spectral index of $-1$ or even harder for electrons is natural in the electrons injected by PWNe, where magnetic reconnection can play an important role \citep{lu21}. Thus we argue the extended VHE emission we found is associated with the pulsar rather than some unknown hadronic accelerators.

Assuming a distance of 2~\rm kpc, the total energy required in the electrons to account for the emission is less than 1 percent of the total spin-down energy ($L_{\rm sd}\times\tau$) released by PSR J0248+6021. 
Given the morphology of the detected VHE \gray emission, a natural explanation is that this LHAASO source is the TeV pulsar halo or the PWN associated with PSR J0248+6021.  We discussed both scenarios in the following.  

As shown in Sec.\ref{sec:lhaaso}, for both KM2A and WCDA we found a slight improvement in likelihood fitting using the diffusion model rather than a simple Gaussian model. The diffusion model is predicted for pulsar halos, in which the electron propagation is 'free' diffusion in the ISM.  Also in the pulsar halo scenario, the diffusion coefficient is about $2\times10^{28}(d/2.0\,\rm{kpc})^{2}\,\rm{cm}^2/s$ for electrons/positrons at$\sim$ 160~\rm TeV and with a cooling time of $\sim$5.5~\rm kyr assuming a magnetic field strength of 3\,$\mu$G and a CMB energy density of 0.26\,eV\,cm$^{-3}$, which is significantly higher than the diffusion coefficient in Geminga and PSR J0622+3749, but not far from the value in Monogem  (see Tab.~\ref{table:pulsar1} and Tab.~\ref{table:pulsar2} for comparison). In this scenario, we derived the parameter $\theta_d$ which is related to the diffusion length $l_d$ by $\theta_d \sim l_d/ d$, and $ d$ is the distance. In the diffusion regime, $l_d \sim\sqrt{Dt_{\rm cool}}$, where $D=D_0 E^{\delta}$ is the diffusion coefficient, $t_{\rm cool} \sim E^{-1}$ is the electron cooling time, and $E$ is the energy of electrons.  When $\frac{4\gamma kT}{m_ec^{2}} >>1$, where $\gamma$ is the Lorentz factor, $k$ is the Boltzmann constant, $m_e$ is the rest mass of the electron, which corresponds to high-energy scenarios where the photon energy becomes comparable to or larger than the electron rest mass energy, the IC scattering process is in Klein-Nishina (KN) regime. For IC scattering of electrons with CMB photons, the KN regime is achieved when $E >$200$~\rm$TeV, and the cooling time scale follows $t_{\rm cool} \sim E^{-0.7}$ \citep{Khangulyan_2014}. As a result, the energy dependence of $\theta_d$ can be described as $\theta_d \sim E_{<200~\rm TeV}^{(\delta-1)/2}$ ($\theta_d \sim E_{>200~\rm TeV}^{(\delta-0.7)/2}$). Assuming a Kolmogolov-type turbulence spectrum, $\delta=1/3$,  $\theta_d \sim E_{<200~\rm TeV}^{-1/3}$  ($\theta_d \sim E_{>200 ~\rm TeV}^{-0.183}$), which predicted a decrease of the halo size with energy. The electron energy $E$ is related with the \gray energy $E_{\gamma}$ as $\frac{E_{\gamma}}{E} = \frac{t}{t+0.3} \frac{log(1+t/4)}{log(1+t)}$, where $t = 4EkT$, $k$ is Boltzmann constant, assuming IC scattering of electrons with CMB \citep{Khangulyan_2014}. In the energy range of interests here ($[1-200]~\rm TeV$), $E = 20.26\times E_{\gamma}^{0.049\times log(E_\gamma)+0.530}$ can fit the relation with an accuracy of 5$\%$ .
As a result,  the energy dependence of the halo morphology gives us a direct measurement of the energy dependence of the diffusion coefficient and thus the property of the magnetic turbulence in the ISM. In this regard, the large energy band covered by WCDA and KM2A provides us the first chance to perform such kind of study. As shown in Tab.~\ref{tab:tab1}, $\theta_d$ are measured by WCDA and KM2A separately. Due to the high significance of KM2A detection we further divided the KM2A results into two energy bins of $[25,63]$ TeV and $[63,251]$ TeV. The results are shown in Fig.~\ref{fig:Figure2}.  We found that, due to the large statistical errors,  the current data are consistent with  Bohm diffusion ($\delta=1$, thus $\theta_d \sim E_{< 200 ~\rm TeV}^0$, $\theta_d \sim E_{>200 ~\rm TeV}^{0.15}$),  Kolmogorov-type diffusion and energy-independent diffusion ($\delta=0$, thus $\theta_d \sim E_{< 200 ~\rm TeV}^{-0.5}$, $\theta_d \sim E_{>200 ~\rm TeV}^{-0.35}$). In addition, the proper motion of the pulsar can also induce a larger extension in the lower energy.  We note that in fact the the energy dependence of $\theta_d$ and thus the measurement of and  $\delta$ is influenced not only by the morphology of the source but also by the shape of the parent electrons, although we do not consider the shape of the SED here.

Another interesting prediction of the pulsar halo scenario is that due to the proper motion of the pulsar, the electrons injected at the different stages of the pulsar evolution will have different spatial distributions. The electrons injected earlier would concentrate near the birthplace of the pulsar. And due to effective cooling these "old" electrons can now only emit GeV $\gamma$-rays. Di Mauro et al.\citep{dimauroDetectionRayHalo2019} has found hints for such GeV structures for Geminga. In Fig.~\ref{fig:Figure4} we showed the Fermi LAT TS map above $1~ \rm GeV$. The birth position of the pulsar PSR J0248+6051 is also shown assuming the age of 65 kyrs and the proper motion velocity $\sim 500~\rm km/s$ \citep{Theureau2011}.  We found no significant GeV emission in this region.  However, we note that the PSR J0248+6051 itself is extremely bright in the GeV band and may hinder any extended emissions in this energy range. 

 With the current data, we didn't find a decisive preference of the diffusion model and thus the observed \gray emission can also be interpreted as a PWN.  The derived physical size of the Gaussian template assumption is of the order of $10~\rm pc$, which is possible for old  PWNe with an age of dozens of thousand years \citep{mitchell22}. Indeed, the age of PSR J0248+6021 is significantly younger than other pulsar halos. The no detection of X-ray counterpart of the PWN can be interpreted as that the PWN itself already expanded to fill a larger volume, the average magnetic field has already decreased and the structures have become rather diffuse and dim in X-ray band. The age of about $60,000$ years and the very energetic nature of this pulsar make it a very ideal candidate to study the transition between the 'canonical' PWN to pulsar halos. In the current understanding of PWN evolution, when PWN is crushed by the reverse shock but the pulsar is still within the SNR,  due to the declining magnetic field,  PWN can be observed as an X-ray-dim, \gray-bright ‘relic’ bubbles \citep{lopez-coto22}. Such phase is predicted for pulsars with ages of $10^4$ to $10^5$ yrs, which is consistent with the age of PSR J0248+6021.

In both scenarios mentioned above, an extended X-ray structure due to the synchrotron radiations from the same population of VHE electrons is expected and may be hindered in the former observations due to the limited FOV of current X-ray instruments. E-Rosita \citep{erosita} and Einstein Probe \citep{ep} can be the ideal instrument to detect such X-ray structures. The unprecedented angular resolution in X-ray band will provide us with decisive information on this source and ISM properties in this region. Considering the similar spectral shape, the unidentified source LHAASO J0341+5258 and LHAASO J2108+5157 could also potentially be pulsar halos or evolved PWNe, albeit lacking in the reported powerful pulsars \citep{LHAASO_J0341,LHAASO_J2108}.  In this regard, the accumulation of exposure of LHAASO on this source, as well as observations by instruments with better angular resolutions, such as CTA \citep{cta} and LACT \citep{lact} in VHE \gray band, would be extremely crucial in understanding the nature of this system. 

In conclusion, we found extended  TeV \gray emission from both WCDA and KM2A surrounding the energetic PSR J0248+6021. The spectral and morphology of the \gray emission is well fitted in the context of a new pulsar halo, or an evolved extended PWN.  This system is an ideal site to study the transition of PWN to Pulsar halo, as well as the ISM properties. Future high-resolution observations will shed light on these issues. 


\begin{table*}
\centering
\caption{Formulae for parent particle distribution using in the SED fitting process}
\begin{tabular}{ccc}
\hline
\hline
Name & Formula  &Free parameters   \\
\hline  
ECPL & $N(E) = A (E/10\,{\rm TeV})^{-\alpha} {\exp(-(E/E_\mathrm{cut})}^{\beta}) $ & $A$, $\alpha$, $\beta$, $ E_\mathrm{cut}$ \\
BPL &  $N(E) =\begin{cases} A(E/10\,{\rm TeV})^{-\alpha_1} & \mbox{: }E<E_\mathrm{b} \\ A(E_\mathrm{b}/10\,{\rm TeV})^{(\alpha_2-\alpha_1)}(E/10\,{\rm  TeV})^{-\alpha_2} & \mbox{: }E>E_\mathrm{b} \end{cases}$&  $A$, $\alpha_1$, $\alpha_2$, $E_\mathrm{b}$\\
\hline
\end{tabular}
\label{tab:form}
\end{table*}

\begin{table*} 
\centering
	\caption{Best-fit results of different radiation models described in Sec.\ref{sec:dis} }
	\begin{tabular}{ccccccc}
	\hline
Model & Distribution &$\alpha$/$\alpha_1$ &$\beta$ / $\alpha_2$  & $E_{\rm cut}$/$E_{\rm b}$ (TeV)  & MLL $^{a}$\\
	 	\hline
\multirow{2}{*}{PP$^{b}$} &ECPL& $<1$  & 1.07$^{+0.44}_{-0.28}$  &66.24$^{+46.94}_{-36.49}$  & -5.22 \\ 
    & BPL &$<1$ &4.04$^{+0.54}_{-0.41}$ &81.34$^{+19.44}_{-17.11}$ &  -3.62 \\ %
    \hline
    \multirow{2}{*}{IC$^c$ }&ECPL& 0.95$^{+0.47}_{-0.68}$  &0.95$^{+0.29}_{-0.22}$  & 24.81$^{+18.28}_{-14.21}$ & -4.45\\
	& BPL & $<1$ &  4.48$^{+0.26}_{-0.23}$ &44.00$^{+6.27}_{-6.26}$ & -2.45 \\
	\hline 
\end{tabular}
\\
{\footnotesize 
$^a$ Maximum of Log(likelihood).\\
$^b$  $W_p(>1 {\rm GeV})\sim (6/n_{\rm H})\times10^{48}$\,erg, in which $n_{\rm H}$ (in the unit of ${\rm cm}^{-3}$)is the number density of the atomic hydrogen in the medium.\\
$^c$ Using cosmic background emission (CMB) as photon seeds. $W_e(>1 {\rm GeV})\sim 5\times10^{45}$ erg for BPL model.\\

}
\label{tab:sedfit} 
\end{table*}

\begin{table*} 
\centering
\caption {Comparison of the properties of pulsars J0622+3749, 
Geminga, and Monogem.}
\begin{tabular}{ccccccc}
\hline 
Name & $P$ & $\dot{P}$ & $L_{\rm sd}$ & $\tau$ & $d$ & Ref.\\
     & (s) & ($10^{-14}$~s~s$^{-1}$) & ($10^{34}$~erg s$^{-1}$) & (kyr) & (kpc)  \\
\hline
J0248+6021 & $0.217$ & $5.509$  & $21.3$ & 62.4  & 2.0 &\citep{Theureau2011}\\
J0622+3749 & $0.333$ & $2.542$ & $2.7$ & 207.8 & 1.60 & \citep{2012ApJ...744..105P}\\
Geminga    & $0.237$ & $1.098$ & $3.3$ & 342.0 & 0.25 & \citep{2005AJ....129.1993M}\\
Monogem    & $0.385$ & $5.499$ & $3.8$ & 110.0 & 0.29 & \citep{2005AJ....129.1993M}\\
\hline
\end{tabular}
\label{table:pulsar1}
\end{table*}

\begin{table*}
\centering
\caption {Comparison of the properties of extended very high energy emission possible associated with pulsars}
\begin{tabular}{cccccc}
\hline 
Name & Gaussian template  &  physical size & diffuse model &physical size& diffusion coefficient \\
     & $\sigma$ (deg) &  gaussian (pc) &$\theta_d$ (deg) & diffuse (pc) & ($10^{27}\,{\rm cm}^{2} {\rm s}^{-1}$)$^c$\\

\hline
J0248+6021 & $0.37\pm0.07$ &  $\sim12.9$  &$1.11\pm0.28$ &$\sim29.0-48.5$&  20.0 (d/2.0~\rm kpc)$^{2}$  \\
J0622+3749$^a$  & $0.40\pm0.07$ &$\sim11.2$ &$0.91\pm0.20$ & $\sim 19.8-31.0$ & 8.9 (d/1.6~\rm kpc)$^{2}$  \\ 
Geminga$^b$   & --- & ---& $5.5\pm0.7$ & $\sim20.9-27.0$ & 3.2 (d/0.25~\rm kpc)$^{2}$  \\
Monogem$^b$   &---  & --- & $4.8\pm0.6$ & $\sim21.2-27.3$  & 15 (d/0.29~\rm kpc)$^{2}$ \\
\hline
\end{tabular}
\\
\footnotesize{ 
$^a$ Data from  Aharonian et al.\citep{2021PhRvL.126x1103A}.\\
$^b$ Data from  Abeysekara et al.\citep{hawc_geminga}.\\}
$^c$  Diffusion coefficient of 100 TeV electrons for Geminga and Monogem. \\ Diffusion coefficient of 160 TeV electrons for LHAASO J0248+6021 and LHAASO J0622+3749.
\label{table:pulsar2}
\end{table*}

\section*{Acknowledgements}
We would like to thank all staff members who work at the LHAASO site above 4400 meter above the sea level year round to maintain the detector and keep the water recycling system, electricity power supply and other components of the experiment operating smoothly. We are grateful to Chengdu Management Committee of Tianfu New Area for the constant financial support for research with LHAASO data. We appreciate the computing and data service support provided by the National High Energy Physics Data Center for the data analysis in this paper. This research work is supported by the following grants: The National Natural Science Foundation of China No.12393854, No.12393851, No.12393852, No.12393853, No.12205314, No.12105301, No.12305120, No.12261160362, No.12105294, No.U1931201, No.12375107, No.12173039, the Department of Science and Technology of Sichuan Province, China No.24NSFSC2319, Project for Young Scientists in Basic Research of Chinese Academy of Sciences No.YSBR-061, and in Thailand by the National Science and Technology Development Agency (NSTDA) and the National Research Council of Thailand (NRCT) under the High-Potential Research Team Grant Program (N42A650868). 

\section*{Data Availability}

The \fermi\ data used in this work is publicly available, which is provided online by the NASA-GSFC Fermi Science Support Center\footnote{\url{ https://fermi.gsfc.nasa.gov/ssc/data/access/lat/}}.



\bibliographystyle{scichina} 
\bibliography{sample631} 

\end{multicols}
\clearpage
Zhen Cao$^{1,2,3}$,
F. Aharonian$^{4,5}$,
Axikegu$^{6}$,
Y.X. Bai$^{1,3}$,
Y.W. Bao$^{7}$,
D. Bastieri$^{8}$,
X.J. Bi$^{1,2,3}$,
Y.J. Bi$^{1,3}$,
W. Bian$^{9}$,
A.V. Bukevich$^{10}$
% ...
(The LHAASO Collaboration)
$^{1}$ Key Laboratory of Particle Astrophysics \& Experimental Physics Division \& Computing Center, Institute of High Energy Physics, Chinese Academy of Sciences, 100049 Beijing, China\\
$^{2}$ University of Chinese Academy of Sciences, 100049 Beijing, China\\
$^{3}$ TIANFU Cosmic Ray Research Center, Chengdu, Sichuan,  China\\
$^{4}$ Dublin Institute for Advanced Studies, 31 Fitzwilliam Place, 2 Dublin, Ireland \\
$^{5}$ Max-Planck-Institut for Nuclear Physics, P.O. Box 103980, 69029  Heidelberg, Germany\\
$^{6}$ School of Physical Science and Technology \&  School of Information Science and Technology, Southwest Jiaotong University, 610031 Chengdu, Sichuan, China\\
$^{7}$ School of Astronomy and Space Science, Nanjing University, 210023 Nanjing, Jiangsu, China\\
$^{8}$ Center for Astrophysics, Guangzhou University, 510006 Guangzhou, Guangdong, China\\
$^{9}$ Tsung-Dao Lee Institute \& School of Physics and Astronomy, Shanghai Jiao Tong University, 200240 Shanghai, China\\
$^{10}$ Institute for Nuclear Research of Russian Academy of Sciences, 117312 Moscow, Russia\\
% ...

\end{document}